\documentclass[twocolumn,times,10pt]{article}
\usepackage[a4paper]{geometry}
\usepackage{newtxtext,newtxmath}
\usepackage{graphicx}
\usepackage{xcolor}
\usepackage{caption}
\usepackage{subcaption}
\usepackage[per-mode=repeated-symbol]{siunitx}
\usepackage{glossaries}
\usepackage[hidelinks]{hyperref}
\usepackage[noabbrev,capitalise]{cleveref}
\usepackage{booktabs}
\usepackage{array}
\usepackage[symbol]{footmisc}
\usepackage{threeparttable}
\usepackage{makecell}
\usepackage{arydshln}
\usepackage{tikz}
\usepackage{eso-pic}
\usepackage{hyphenat}
\usepackage{microtype}

\newcommand*\samethanks[1][\value{footnote}]{\footnotemark[#1]}
\newcommand*\sqmm{\si{\milli\metre\squared}}

\DeclareSIUnit\flop{FLOP}
\DeclareSIUnit\comp{COMP}

\Crefname{equation}{Eq.}{Eqs.}
\Crefname{figure}{Fig.}{Figs.}
\Crefname{tabular}{Tab.}{Tabs.}

\newcommand{\ssymbol}[1]{^{\@fnsymbol{#1}}}

\ifx\showcolor\undefined
    \newcommand{\todo}[1]{{#1}}
    \newcommand{\dk}{-}
    \newcommand{\vf}[1]{{#1}}
    \ifx\shownum\undefined
        \newcommand{\bs}[1]{{#1}}
        \newcommand{\sh}[1]{{#1}}
    \else
        \newcommand{\bs}[1]{\textcolor{orange}{#1}}
        \newcommand{\sh}[1]{\textcolor{blue}{#1}}
    \fi
    \newcommand{\lb}[1]{{#1}}
    \ifx\showref\undefined
        \newcommand{\ct}[1]{~{[#1]}}
        \newcommand{\rc}[1]{{[#1]}}
        \newcommand{\tn}[1]{~\tnote{#1}}
        \newcommand{\im}[1]{\item[#1]}
        \newcommand{\fg}[1]{#1}
    \else
        \newcommand{\ct}[1]{~{\textcolor{cyan}{[#1]}}}
        \newcommand{\rc}[1]{{\textcolor{cyan}{[#1]}}}
        \newcommand{\tn}[1]{~\textcolor{cyan}{\tnote{#1}}}
        \newcommand{\im}[1]{\item[\textcolor{cyan}{#1}]}
        \newcommand{\fg}[1]{\textcolor{cyan}{#1}}
    \fi
    \newcommand{\lc}[1]{{#1}}
    \newcommand{\ld}[1]{{#1}}
    \ifx\lucalast\undefined
        \newcommand{\lf}[1]{{#1}}
    \else
        \newcommand{\lf}[1]{\textcolor{pink}{#1}}
    \fi
\else
    \newcommand{\todo}[1]{{\textcolor{red}{#1}}}
    \newcommand{\dk}{\todo{?}}
    \newcommand{\bs}[1]{\textcolor{orange}{#1}}
    \newcommand{\sh}[1]{\textcolor{blue}{#1}}
    \newcommand{\lb}[1]{\textcolor{purple}{#1}}
    \newcommand{\vf}[1]{\textcolor{teal}{#1}}
    \newcommand{\ct}[1]{~{\textcolor{cyan}{[#1]}}}
    \newcommand{\rc}[1]{{\textcolor{cyan}{[#1]}}}
    \newcommand{\tn}[1]{~\textcolor{cyan}{\tnote{#1}}}
    \newcommand{\im}[1]{\item[\textcolor{cyan}{#1}]}
    \newcommand{\fg}[1]{\textcolor{cyan}{#1}}
    \newcommand{\lc}[1]{\textcolor{violet}{#1}}
    \newcommand{\ld}[1]{\textcolor{magenta}{#1}}
    \newcommand{\lf}[1]{\textcolor{pink}{#1}}
\fi

\ifx\blind\undefined
    \def\occamy{Occamy}
\else
    \def\occamy{Ogopogo}
\fi

\newacronym{su}{SU}{streaming unit}
\newacronym{la}{LA}{linear algebra}
\newacronym{spm}{SPM}{scratchpad memory}
\newacronym{gp}{GP}{general-purpose}

\newacronym{ieee}{IEEE}{IEEE}
\glsunset{ieee}

\newcommand{\x}{$\times$}
\newcommand{\sep}{$\vert$}

\def\titleoffset{-8.5mm}

\advance\topmargin-1in\advance\oddsidemargin-1in
\evensidemargin\oddsidemargin	     

\makeatletter

\def\@startsection#1#2#3#4#5#6{\if@noskipsec \leavevmode \fi
   \par \@tempskipa #4\relax
   \@afterindenttrue
   \ifdim \@tempskipa <\z@ \@tempskipa -\@tempskipa \@afterindentfalse\fi
   \if@nobreak \everypar{}\else
     \addpenalty{\@secpenalty}\addvspace{\@tempskipa}\fi \@ifstar
     {\@dblarg{\@sect{#1}{#2}{#3}{#4}{#5}{#6}}}%
     {\@dblarg{\@sect{#1}{#2}{#3}{#4}{#5}{#6}}}}

\def\section{\@startsection {section}{1}{\z@}{0pt plus 0pt minus 0pt}
{1pt plus 0pt minus 0pt}{\centering\normalsize\bf}}

\def\subsection{\@startsection {subsection}{2}{\z@}{2pt plus 2pt minus 2pt}
{1pt plus 0pt minus 0pt}{\normalsize\it}}

\long\def\@makecaption#1#2{
\vskip10pt\begin{center}\small #1 #2 \end{center}\par\vskip 1pt}
\def\fnum@figure{\raggedright{\small Fig. \thefigure }.\small}
\def\fnum@table{\raggedright{\small Table \thetable }.\small}

\begin{document}

\AddToShipoutPictureBG*{%
  \AtPageUpperLeft{%
    \hspace{\paperwidth}%
    \raisebox{-\baselineskip}{%
      \makebox[-35pt][r]{\footnotesize{
        \copyright~2024~IEEE. Personal use of this material is permitted. %
        Permission from IEEE must be obtained for all other uses, in any current or future media, including
      }}
}}}%

\AddToShipoutPictureBG*{%
  \AtPageUpperLeft{%
    \hspace{\paperwidth}%
    \raisebox{-2\baselineskip}{%
      \makebox[-37pt][r]{\footnotesize{
        reprinting/republishing this material for advertising or promotional purposes, creating new collective works, for resale or redistribution to servers or lists, or
      }}
}}}%

\AddToShipoutPictureBG*{%
  \AtPageUpperLeft{%
    \hspace{\paperwidth}%
    \raisebox{-3\baselineskip}{%
      \makebox[-185pt][r]{\footnotesize{
       reuse of any copyrighted component of this work in other works.
      }}
}}}%

\date{}

\title{\Large\bf
\vspace{\titleoffset}
{\occamy}: A \vf{432}-Core \sh{28.1} DP-GFLOP/s/W \sh{83}\% FPU Utilization Dual-Chiplet, Dual-HBM2E RISC-V-based Accelerator for Stencil and Sparse Linear Algebra \lc{Computations} with \vf{8-to-64}-bit Floating-Point Support \lb{in \vf{12nm FinFET}}\vspace{-0.6cm}}

\ifx\blind\undefined
\author{\large%
Gianna Paulin,\textsuperscript{\!\footnotesize*}
Paul Scheffler,\textsuperscript{\!\!\footnotesize*%
    \stepcounter{footnote}%
    \stepcounter{footnote}%
    \stepcounter{footnote}%
    \thanks{ Authors contributed equally to this work.}\,%
}
Thomas Benz,\textsuperscript{\!\!\footnotesize*\samethanks}\,
Matheus Cavalcante,\textsuperscript{\!\footnotesize\textdagger}
Tim Fischer,\textsuperscript{\!\footnotesize*}
Manuel Eggimann,\textsuperscript{\!\footnotesize*} \\
Yichao Zhang,\textsuperscript{\!\footnotesize*} 
Nils Wistoff,\textsuperscript{\!\footnotesize*}
Luca Bertaccini,\textsuperscript{\!\footnotesize*}
Luca Colagrande,\textsuperscript{\!\footnotesize*}
Gianmarco Ottavi,\textsuperscript{\!\textdaggerdbl} \\
Frank K. G\"{u}rkaynak,\textsuperscript{\!\footnotesize*}
Davide Rossi,\textsuperscript{\!\textdaggerdbl}
Luca Benini\textsuperscript{\footnotesize*\textdaggerdbl} \\
\normalsize  \textsuperscript{\footnotesize*}ETH Zurich, Switzerland\hspace{1.5cm} \textsuperscript{\footnotesize\textdagger}Stanford University, USA\hspace{1.5cm} \textsuperscript{\footnotesize\textdaggerdbl} University of Bologna, Italy \\
\vspace{-1.30cm}
}
\else
\author{\centering{\textit{Authors omitted for blind review.}}}
\fi

\maketitle
\thispagestyle{empty}

\noindent\textbf{Abstract:} We present {\occamy}, a \vf{432}-core RISC-V dual\hyp chiplet 2.5D system for efficient sparse linear algebra and stencil computations on FP64 and narrow (\mbox{32-,} \mbox{16-,} 8-bit) SIMD FP data. {\occamy} features \vf{48} clusters of RISC-V cores with custom extensions, \lb{\vf{two} 64-bit host cores, and a latency-tolerant \lc{multi-chiplet interconnect and} memory system with \SI{32}{\gibi\byte} of HBM2E}. %
\ld{It achieves leading-edge utilization on stencils (\sh{\SI{83}{\percent}}), sparse-dense (\sh{\SI{42}{\percent}}), and sparse-sparse (\sh{\SI{49}{\percent}}) matrix multiply.}


\section{Introduction}
\noindent Sparse machine learning (ML) and high-performance computing applications in fields like \lb{multiphysics} simulation and graph analytics often rely on sparse \gls{la}, stencil codes, and graph pattern matching\ct{1}.
These workloads \lb{achieve low FPU utilization (typically $<\sh{\SI{10}{\percent}}$ for sparse \gls{la})} \lc{on modern CPUs and GPUs because of their sparse, irregular memory accesses \ld{and} complex\ld{, indirection-based} address computation\ld{s}\ct{2-4}.}
\lc{While many specialized accelerators have been proposed for sparse ML workloads, they lack \ld{the flexibility of instruction processors}\ct{5}.}
We present {\occamy}, \lc{a flexible, general-purpose}, dual-chiplet system with \vf{two} \SI{16}{\gibi\byte} HBM2E stacks optimized for \lc{a wide range} of irregular-memory-access workloads.
Each chiplet \lb{integrates} a RISC-V \lc{host} core and \vf{216} lightweight, latency-tolerant RISC-V compute cores with domain-specific ISA extensions organized hierarchically in \vf{six} groups of \vf{four} \lb{\vf{nine}-core} compute clusters.
{\occamy} \lf{demonstrates in silicon three innovations}: \textbf{(A)} efficient multi-precision compute cores with \lb{sparse} \glspl{su} supporting indirection, intersection, and union operations to accelerate general sparse \lb{computations}, \textbf{(B)} a scalable, latency-tolerant, hierarchical architecture with separate data and control interconnects and distributed DMA units for agile on-die and die-to-die traffic, and \textbf{(C)} \lb{an innovative} system-in-package 2.5D integration for two compute chiplets with two \SI{16}{\gibi\byte} HBM2E stacks.

\section{Architecture}
\noindent Within each cluster, \vf{eight} worker cores and one DMA control core share \SI{128}{\kibi\byte} of tightly coupled \gls{spm} through a single-cycle interconnect (\Cref{fig:arch}\fg{f}).
Each worker core features a 64-bit-wide SIMD FPU supporting FP64, FP32, FP16, FP16alt (8,7), FP8, and FP8alt (4,3).
In addition to the typical fused multiply-accumulate \lb{(FMA)} instructions, the FPUs support widening sum-dot-product and three-addend summation instructions for FP8 and FP16 formats\ct{6}.
Two worker-core ISA extensions maximize the FPU utilization for both regular and irregular workloads: a hardware loop buffer~(\Cref{fig:arch}\fg{h}) and three register-interfaced sparsity-capable \glspl{su}\ct{7}~(\Cref{fig:sssr}).
Each \gls{su} may read or write the SPM using $\leq$4D affine decoupled streams to accelerate regular \lb{tensor} workloads.
\lb{Notably, two \glspl{su}} support indirect (8-, 16-, or 32-bit-indexed) streams to accelerate sparse-dense \gls{la} and stencil codes, as shown in \Cref{fig:code}, and can cooperate to intersect or merge sparse tensors \lc{in} sparse-sparse LA or graph matching.
The third \gls{su} can write out joint indices for sparse result tensors.
Each group of four clusters (\Cref{fig:arch}\fg{e}) shares \vf{\SI{64}{\gibi\byte\per\second}} of bandwidth to the local HBM2E stack and other groups.
\Cref{fig:arch}\fg{d} shows the chiplet top level.
Each cluster's DMA core supports 1D and 2D transfers to move data \lb{blocks} over the latency-tolerant hierarchical 512-bit data interconnect\ct{8}.
The 64-bit host manages the groups and peripherals over a secondary 64-bit interconnect. 
The chiplets communicate over two all-digital source-synchronous \lb{fault-tolerant} double-data-rate die-to-die links (D2Ds) (\Cref{fig:arch}\fg{b}): a 64- and a 512-bit D2D with one and \vf{38} duplex physical channels, respectively.

\section{Results}
\begin{figure}[ph!]

\vspace{0.1cm}
\centering
\begin{minipage}{\textwidth}
\centering
\includegraphics[width=\linewidth]{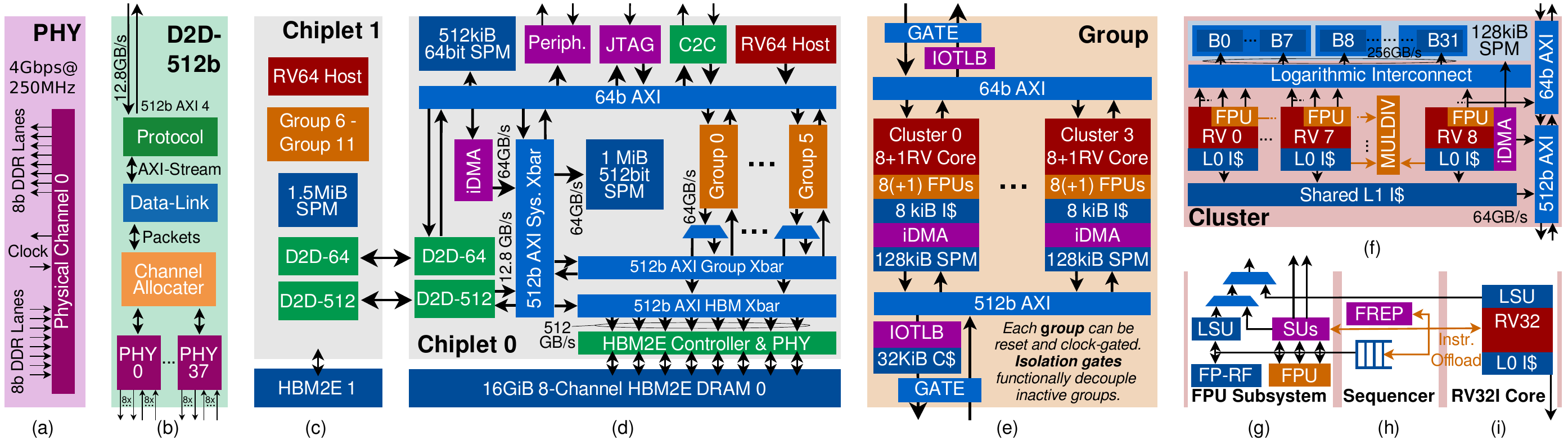}
\vspace{-1.0cm}
\caption{Architecture of our dual-chiplet system. (c, d): chiplets connected by a 512-bit D2D link (a, b). One chiplet (c, d) contains six groups, peripherals, a 64-bit Linux-capable host and 1.5MiB SPM. One group (e) contains four clusters (f) containing eight compute coreplexes (g-i).}
\label{fig:arch}
\end{minipage}%

\vspace{-0.3cm}
\centering
\begin{minipage}{.36\textwidth}
  \centering
  \includegraphics[width=\linewidth]{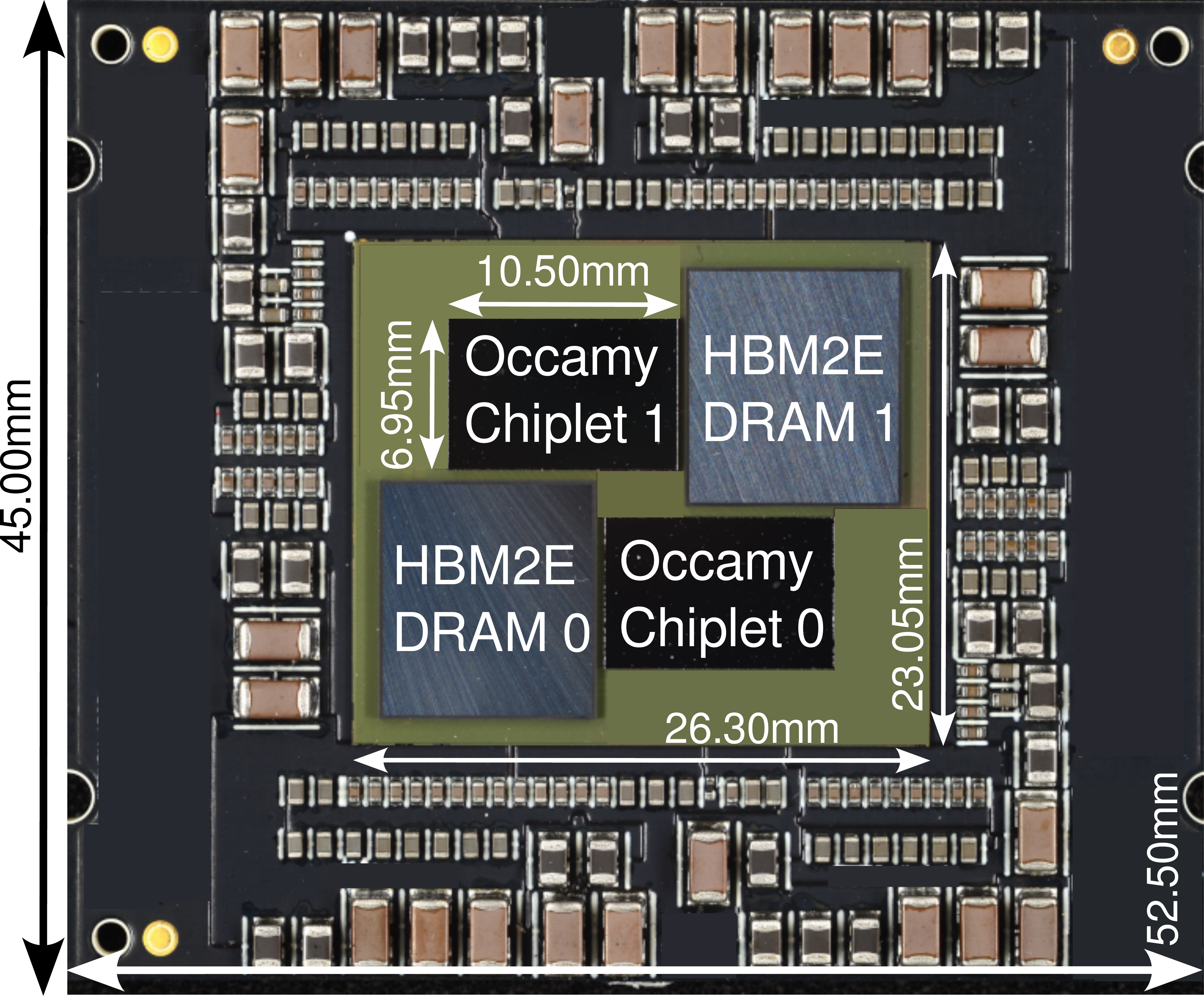}
  \vspace{-0.8cm}
  \captionof{figure}{Module photograph with dimensions.}
  \label{fig:dieshot}
\end{minipage}%
\quad
\begin{minipage}{0.60\textwidth}
\begin{minipage}{\textwidth}
  \centering
  \includegraphics[width=0.85\linewidth]{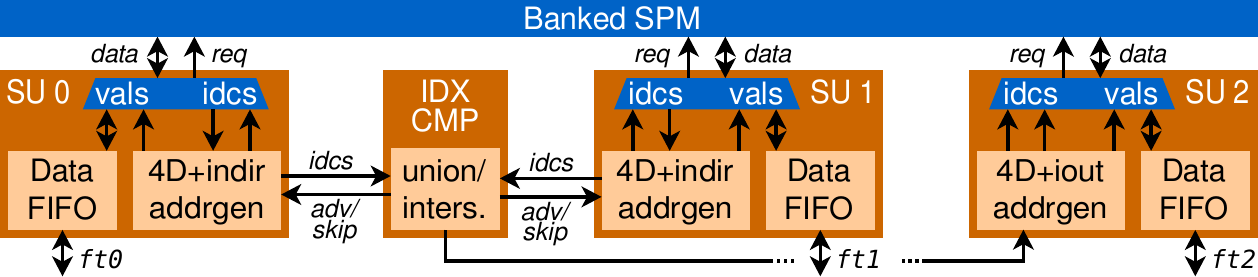}
  \vspace{-0.4cm}
  \captionof{figure}{Architecture of the cooperating sparsity-capable SUs in each worker core.}
  \label{fig:sssr}
\end{minipage}%

\vspace{-0.25cm}
\begin{minipage}{\textwidth}
  \centering
  \includegraphics[width=0.97\linewidth]{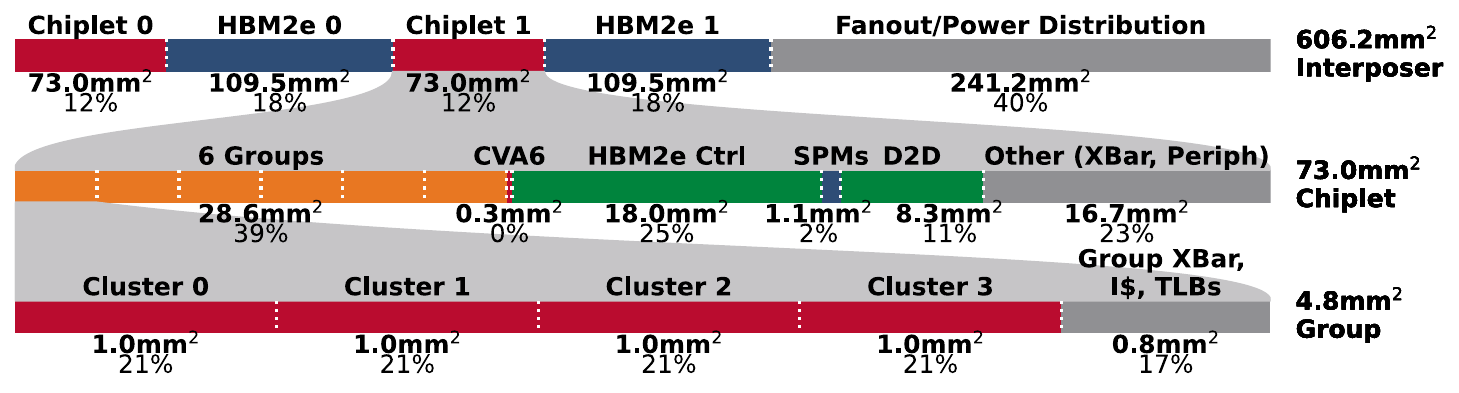}
  \vspace{-0.55cm}
  \captionof{figure}{Hierarchical area breakdown of interposer and chiplets.}
  \label{fig:area_plot}
\end{minipage}
\end{minipage}

\vspace{-0.28cm}
\centering
\begin{minipage}{0.26\textwidth}
  \centering
  \includegraphics[width=\linewidth]{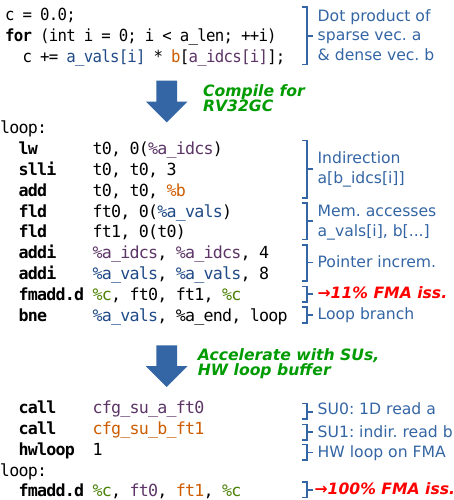}
  \vspace{-0.8cm}
  \captionof{figure}{Sparse-dense dot product assembly without and with our SUs\ld{; our ISA extensions enable} continuous FMA \lf{execution}.}
  \label{fig:code}

\end{minipage}%
\quad
\begin{minipage}{0.72\textwidth}
  \centering
  \includegraphics[width=\linewidth]{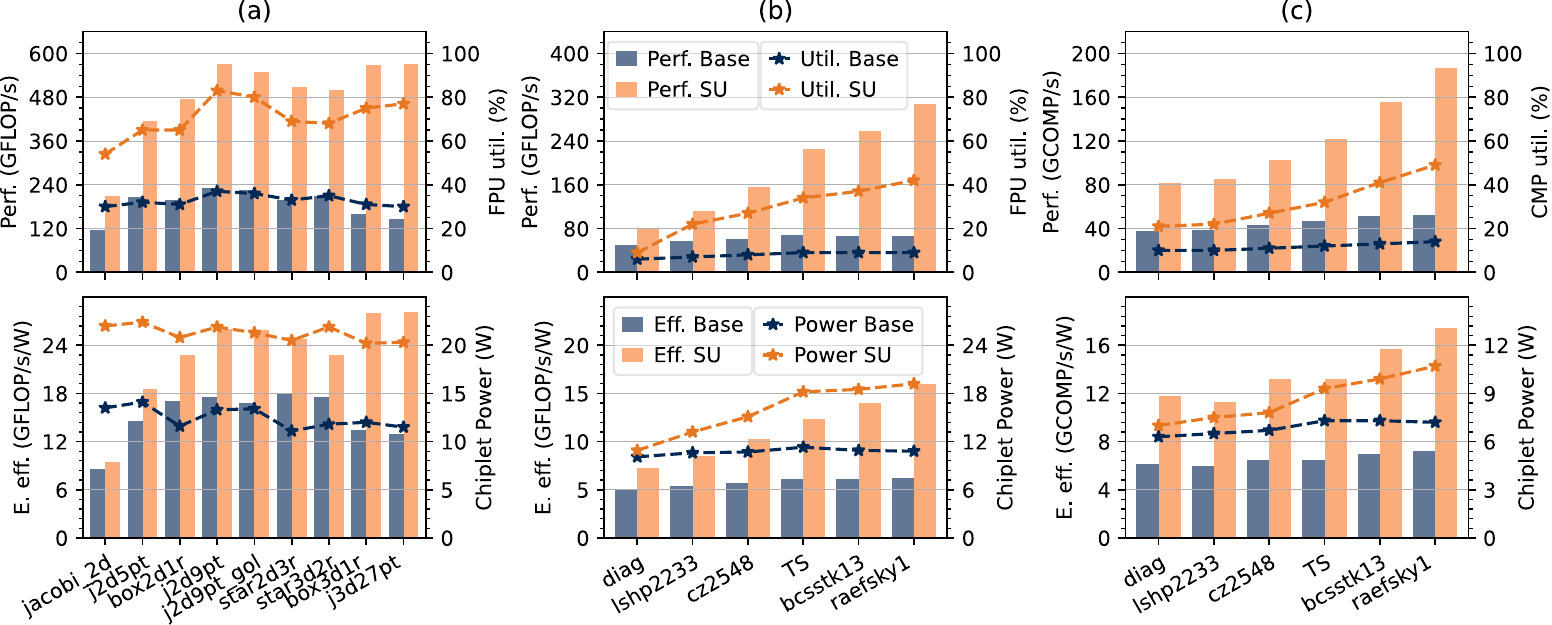}
  \vspace{-0.9cm}
  \captionof{figure}{Performance (top), energy efficiency and power (bottom) for FP64 stencil codes (a), SpMM (b) and SpMSpM (c), including the impact of on-chiplet double-buffered transfers. \ld{SpM(Sp)M} left matrices (X axis) are real-world \ld{sparse} matrices.  SpMSpM right matrices are random with \SI{1}{\percent} density.}
  \label{fig:perf}
\end{minipage}%

\vspace{-0.43cm}
\centering
\begin{minipage}{.221\textwidth}
  \centering
  \vspace{0.25cm}
  \includegraphics[width=\linewidth]{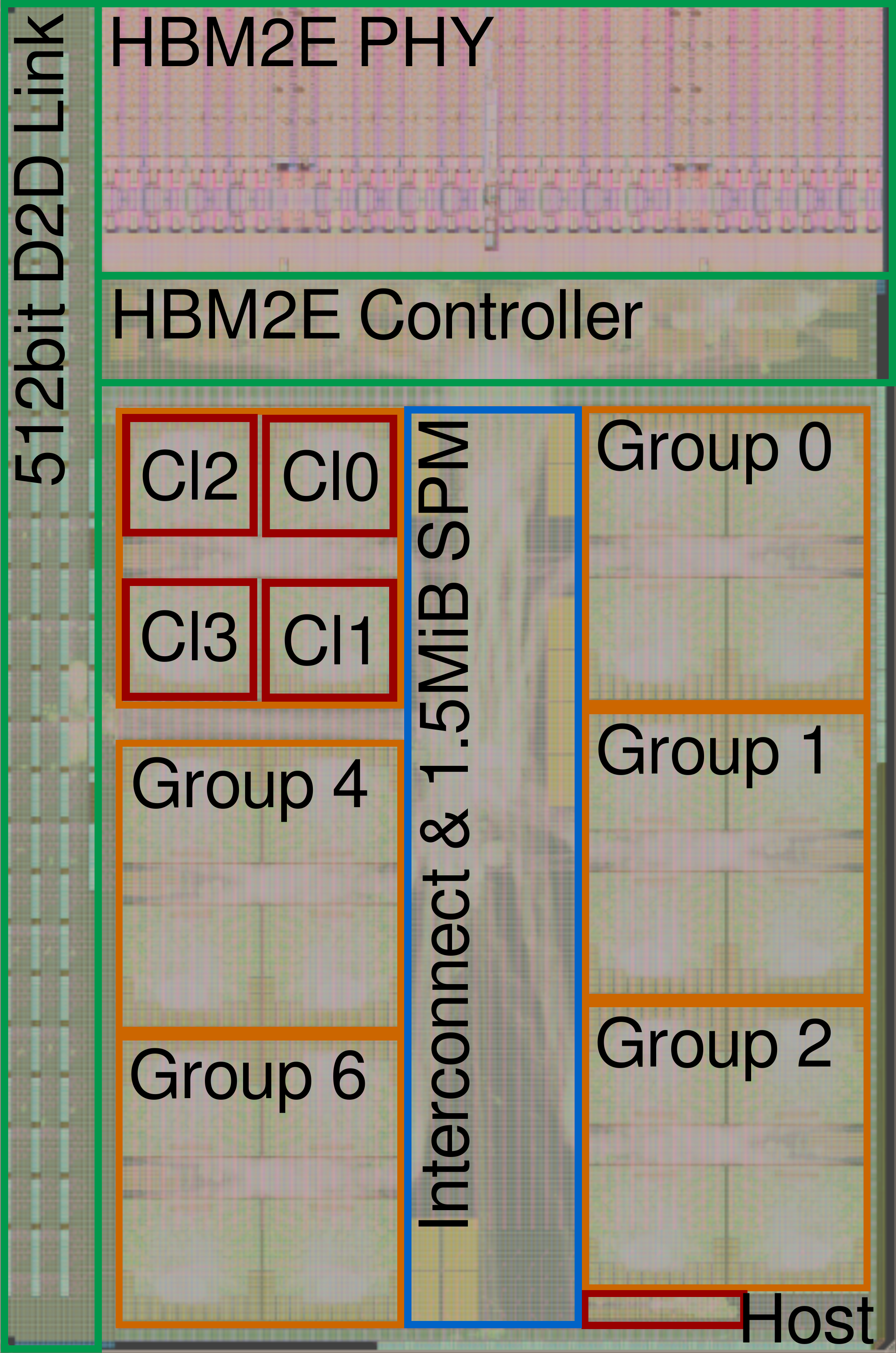}
  \vspace{-0.75cm}
  \captionof{figure}{Chiplet floorplan.}
  \label{fig:floorplan}
\end{minipage}%
\hspace{0.025cm}
\begin{minipage}{0.76\textwidth}%
    \centering
    \fontsize{6.8pt}{6.8pt}\selectfont
    \renewcommand{\arraystretch}{0.9}
    \setlength{\tabcolsep}{0.12cm}
    \begin{threeparttable}
\begin{tabular}{@{}llllll@{}}
\toprule
&
\textbf{Schmidt et al.}\ct{9} & %
\textbf{Fujitsu A64FX}\ct{10} & %
\textbf{AMD Rome}\ct{11-12} & %
\textbf{Nvidia A100}\ct{13} & %
\textbf{\textit{{\occamy} [Ours]}} \\
\midrule
Package &
Die on PCB &
2.5D Passive Interp. &
2.5D Interposer &
2.5D Interposer &
2.5D Passive Interp. \\
\hdashline 
Configuration &
1 Die &
1 Die, 4 HBMs &
8 Dies, 1 IO Die &
1 Die, 6 HBMs &
2 Dies, 2 HBMs \\
\hdashline 
Die technology &
16nm FinFET &
7nm FinFET &
7nm FinFET &
7nm FinFET &
12nm FinFET\tn{a} \\
\hdashline 
Transistors &
\vf{0.5B} &
\vf{8.8B} &
\bs{67B} &
\vf{54B} &
\bs{4.8B} \\
\hdashline 
\makecell[cl]{Die Area [\sqmm] \\ Compute Area [\sqmm]\tn{b}} &
\makecell[cl]{\vf{24.0} \\ \sh{3.86} (\sh{0.48}/core)} &
\makecell[cl]{\vf{420} \\ \sh{243} (\sh{5.1}/core)} &
\makecell[cl]{\bs{592}\tn{c} \\ \sh{442} (\sh{6.9}/core)} &
\makecell[cl]{\vf{826} \\ \sh{632} (\sh{5.9}/SM)} &
\makecell[cl]{\vf{146} \\ \sh{51.5} (\vf{1.0}/cluster)} \\
\hdashline
System DRAM &
\vf{0} &
\vf{32GiB} HBM2 &
\vf{$\leq$4TiB} DDR4 &
\vf{40GiB} HBM2 &
\vf{32GiB} HBM2E \\
\hdashline
SRAM &
\vf{4.5MiB} cache &
\vf{32MiB} L2\$ &
\vf{128MiB} L3\$ &
\vf{40MiB} L2\$ &
\bs{9MiB} SPM \\
\hdashline
ISA &
RV64GXhwacha4 &
Armv8.2-A SVE\tn{d} &
x86-64, AVX2 &
PTX ISA, CC 8.0 &
RV32GXoc\tn{e} \tn{f} \\
\hdashline
\makecell[cl]{%
Frequency [\si{\giga\hertz}] (Voltage [V])} &
\vf{1.44} (\vf{1.0}) &
\vf{2.2} (\vf{0.8}) &
\vf{2.3} (\vf{0.8}) &
\vf{1.4} (\vf{0.8}) &
\vf{1.0} (\vf{0.8}) \\
\hdashline
\makecell[cl]{%
Peak FP Perf. \\
$[$TFLOP/s$]$ \\
$[$GFLOP/s/\sqmm$]$ \\
$[$GFLOP/s/\sqmm$]$\tn{b}
} &
\makecell[cl]{%
\emph{FP64/32/16} \\
\vf{0.092/0.18/0.37} \\
\sh{20.9/41.8/83.7}\\
\sh{23.9/47.7/95.4}
} &
\makecell[cl]{%
\emph{FP64/32/16} \\
\vf{3.38/6.76/13.5} \\
\sh{21.9/43.9/87.8} \\
\sh{13.9/27.9/55.7}
} &
\makecell[cl]{%
{FP64/32} \\
\sh{2.3\tn{h}\;\,/4.6\tn{h}} \\
\sh{8.2\tn{h}\;\,/16.4\tn{h}} \\
\sh{5.2\tn{h}\;\,/10.4\tn{h}}
} &
\makecell[cl]{%
\emph{FP64/TF32/TF16} \\
\vf{19.5/312\tn{i}\;\,/624\tn{i}} \\
\sh{48.6/777\tn{i}\;\,/1555\tn{i}} \\
\sh{30.8/493\tn{i}\;\,/987\tn{i}}
} &
\makecell[cl]{%
\emph{FP64/32/16/8} \\
\vf{0.77/1.54/3.07/6.14} \\
\sh{14.9/29.8/59.6/119} \\
\sh{14.9/29.8/59.6/119}
} \\
\hdashline
\makecell[cl]{%
Best FP64 stencil~-~FPU util. \\
$[$GFLOP/s/W$]$ \\
$[$GFLOP/s/\sqmm$]$
} &
\makecell[cl]{%
\dk \\
\dk \\
{\dk}
} &
\makecell[cl]{%
\sh{\SI{11}{\percent}}\tn{g} \ct{14} \\
{\dk} \\
\sh{2.10} {\sep} \sh{1.33}\tn{b} \ct{14}
} &
\makecell[cl]{%
\sh{\SI{37}{\percent}}\tn{h} \ct{15} \\
{\dk} \\
\sh{3.05} {\sep} \sh{1.93}\tn{b} \ct{15}
} &
\makecell[cl]{
\sh{\SI{49}{\percent}}\tn{k} \ct{16} \\
{\dk} \\
\sh{\bf{12.0}} {\sep} \sh{7.59}\tn{b} \ct{16}
} &
\makecell[cl]{%
\sh{\textbf{\SI{83}{\percent}}} \\
\sh{\textbf{28.1}} \\
\sh{11.1}  {\sep} \sh{\textbf{11.1}}\tn{b}
} \\
\hdashline
\makecell[cl]{%
Best FP64 sparse-dense~-~FPU util. \\
$[$GFLOP/s/W$]$ \\
$[$GFLOP/s/\sqmm$]$
} &
\makecell[cl]{%
\dk \\
\dk \\
{\dk}
} &
\makecell[cl]{%
\sh{\SI{4.7}{\percent}}\tn{g} \ct{2} \\
\dk \\
\sh{0.85} {\sep} \sh{0.54}\tn{b} \ct{2}
} &
\makecell[cl]{%
\sh{\SI{8.1}{\percent}}\tn{h} \ct{3} \\
\dk \\
\sh{0.67}{\sep} \sh{0.42}\tn{b} \ct{3}
} &
\makecell[cl]{%
\sh{\SI{2.9}{\percent}}\tn{k} \ct{4} \\
\dk \\
\sh{0.71} {\sep} \sh{0.45}\tn{b} \ct{4}
} &
\makecell[cl]{%
\sh{\textbf{\SI{42}{\percent}}} \\
\sh{\textbf{16.0}} \\
\sh{\textbf{5.95}} {\sep} \sh{\textbf{5.95}}\tn{b}
} \\
\bottomrule
\end{tabular}
\begin{tablenotes}[para, flushleft]
    \fontsize{6.4pt}{6.4pt}\selectfont
    \im{a} Interposer: 65nm
    \im{b} Scaled to 12LP+ node using NAND2X1 gate areas (\si{GE\per\micro\metre\squared})
    \im{c} w/o IO die
    \im{d} 512 bit
    \im{e} Xoc := Xdma\_Xssr\_Xminifloat
    \im{f} Host: RV64GC
    \im{g} At below-peak clock
    \im{h} At base (non-boost) clock
    \im{i} Fully-utilized tensor cores
    \im{k} Tensor cores unused (excl. from FPU util.)
\end{tablenotes}
\end{threeparttable}

    \label{tab:soa}
    \vspace{-0.41cm}
    \captionof{table}{Comparison of our architecture to state-of-the-art.}
\end{minipage}

\end{figure}

\fg{Figs.~7~and~2} show the chiplet floorplan and assembled 2.5D system.
The \vf{\SI{73}{\milli\metre\squared}} chiplets were fabricated in GlobalFoundries' (GF)
\SI{12}{\nano\meter} LP+ node and mounted on a \vf{passive \SI{65}{\nano\meter} PKG-25SI interposer from GF}.
\Cref{fig:perf} summarizes {\occamy}'s performance and energy efficiency based on room-temperature silicon measurements with and without \glspl{su}.
\gls{su} indirection accelerates FP64 stencil codes by up to \sh{3.9$\times$} \lc{compared to assembly-optimized standard RISC-V ISA code}, achieving \sh{\SI{571}{\giga\flop\per\second}}, \sh{\SI{28.1}{\giga\flop\per\second\per\watt}}, and \sh{\SI{83}{\percent}} FPU utilization (\lf{peak performance on \emph{j3d27pt}}).
It also accelerates FP64 sparse-dense matrix multiply (SpMM) by up to \sh{4.6$\times$}, \lc{achieving} up to \sh{\SI{307}{\giga\flop\per\second}}, \sh{\SI{16.0}{\giga\flop\per\second\per\watt}}, and \sh{\SI{42}{\percent}} FPU utilization.
We introduce index comparison rate as a \lf{figure of merit for} sparse-sparse \lf{matrix computation} performance;
our \glspl{su} accelerate FP64 sparse-sparse matrix multiply (SpMSpM) (\sh{\SI{1}{\percent}} right matrix density) by up to \sh{3.6$\times$}, reaching up to \sh{\SI{187}{\giga\comp\per\second}}, \sh{\SI{17.4}{\giga\comp\per\second\per\watt}}, and index comparator utilizations of up to \sh{\SI{49}{\percent}}.
\fg{Tab. 1} compares {\occamy} to state-of-the-art CPUs and GPUs\ct{9-13}, achieving \sh{1.7\x}/\sh{5.2\x} higher FPU utilization and \sh{1.5\x}/\sh{11.0\x} higher \lc{compute-density}- and technology-adjusted performance on stencil/sparse-dense workloads.

\ifx\blind\undefined
{\fontsize{9pt}{9pt}\selectfont
\noindent%
\textls[-23]{
\textbf{Acknowledgment:}  %
We thank F. Zaruba, F. Schuiki, S. Riedel, A. Di Mauro, S. Mach, S. Arjmandpour, N. Huetter, A. Kurth, A. Fontao, B. Muheim, Z. Jiang, G. Rutishauser, S. Scherr, A. Rossi.
We thank GlobalFoundries, Avery, Micron, Rambus, and Synopsys for their generous support.
}
}
\else
\noindent\textbf{Acknowledgment:} \textit{Omitted for blind review.}
\fi

\noindent\textbf{References:}
{\small\noindent 
\rc{1} Z. Zhang et al., \gls{ieee} Access '15;
\rc{2} C. Alappat et al., \gls{ieee}/ACM PMBS '20;
\rc{3} C. Alappat et al., \gls{ieee} TPDS '23
\rc{4} Y. Niu et al. \gls{ieee} IPDPS '21
\rc{5} V. Isaac–Chassande et al., ACM TACO '24;
\rc{6} L. Bertaccini et al., \gls{ieee} ARITH '22;
\rc{7} P. Scheffler et al., \gls{ieee} TPDS '23;
\rc{8} T. Benz et al., \gls{ieee} TCOMP '23;
\rc{9} C. Schmidt et al., \gls{ieee} ISSCC '21;
\rc{10} S. Yamamura et al., \gls{ieee} ISSCC '22;
\rc{11} T. Singh et al. \gls{ieee} ISSCC '20;
\rc{12} S. Naffziger et al. \gls{ieee} ISSCC '20;
\rc{13} J. Choquette et al., \gls{ieee} ISSCC '21;
\rc{14} K. Hirokawa et al., SAGE IJHPCA '22;
\rc{15} L. Szustak et al., \gls{ieee} TPDS '21;
\rc{16} L. Zhang et al., ACM ICS '23;
}

\onecolumn

\end{document}